\documentclass[twocolumn,aps]{revtex4}
\usepackage{epsfig}
\newcommand{\beq}{\begin{eqnarray}}
\newcommand{\eeq}{\end{eqnarray}}
\begin{document}
\title{Mixed Heavy Quark Hybrid Mesons, Decay Puzzles, and RHIC}
\author{Leonard S. Kisslinger\\
Department of Physics, Carnegie Mellon University, Pittsburgh, PA 15213}

\begin{abstract} We estimate the energy of the lowest charmonium and
upsilon states with hybrid admixtures using the method of QCD Sum Rules.
Our results show that the $\Psi'(2S)$ and $\Upsilon(3S)$ states both have 
about a 50\% admixture of hybrid and meson components. From this we
find explanations of both the famous $\rho-\pi$ puzzle for charmonium, and
the unusual pattern of $\sigma$ decays that have been found in $\Upsilon$
decays. Moreover, this picture can be used for predictions of heavy quark
production with the octet model for RHIC.

\end{abstract}
\maketitle
\noindent
PACS Indices:14.40.Gx,12.38.Aw,11.55.Hx,13.25.Gv
\vspace{1mm}

\section{Introduction}

   There is a great interest in studying states with active glue, such as
hybrid mesons, a color singlet composed of a quark-antiquark in a color octet
and a gluon, in order to better understand nonperturbative QCD. 
Recently we have used the method of QCD Sum Rules
in an attempt to find the lowest hybrid charmonium state\cite{kpr08}.
Our conclusion was that the physical states with active glue must be mixed
states, with both charmonium and hybrid charmonium components. In the
present work we use QCD Sum Rules for $J^{PC}=1^{--}$ vector states to
find the lowest mixed meson-hybrid meson states for both charmonium and
upsilon systems.

   In addition to the importance of finding states with active glue, we
are motivated by several experimental considerations. First, the ratio of 
hadronic decays of the charmonium $\Psi'(2S)$ compared to the $J/\Psi(1S)$ 
state is more than an order of magnitude smaller than predicted by 
perturbative QCD (PQCD), the so-called  $\rho-\pi$ puzzle, which was 
discussed at length in Ref\cite{kpr08}. Second, the $\Upsilon(nS)$ states 
have an unusual  pattern of decays into two pions, which also cannot be 
consistent with PQCD\cite{hv06}, which we call the Vogel $\Upsilon(\Delta 
n=2)$ puzzle. Thirdly, our theory of heavy quark states provides a basis 
for the color octet model predictions of RHIC heavy quark 
production\cite{nlc03,cln04}

   In Sec. II we discuss the motivation for the present work on mixed
heavy quark and hybrid mesons. In Sec III we review the method of QCD sum 
rules, the work in Ref\cite{kpr08} on hybrid charmonium, and apply the 
method of QCD Sum Rules for mixed meson-hybrid meson charmonium 
and upsilon states. In Sec. IV we discuss our solution to the $\rho-\pi$ 
puzzle, the Vogel $\Upsilon(\Delta n=2)$ puzzle, and applications of
our mixed hybrid states for the RHIC search for the QCD phase
transition via heavy quark state production. In Sec. V we review our
conclusions.

\section{Heavy Quark Puzzles and RHIC Experiments}

   First let us look at the lowest charmonium and upsilon $(nS)$ states
(FIG 1):

\begin{figure}[ht]
\begin{center}
\epsfig{file=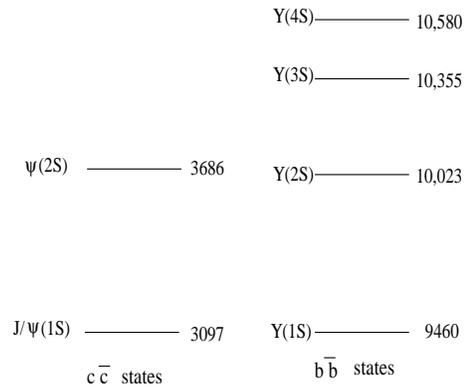,height=5cm,width=6cm}
\caption{Lowest nS charmonium and upsilon states}
\label{Fig.1}
\end{center}
\end{figure}

Note that the separation in energy between the $\psi'(2S)$ and
$J/\psi(1S)$ states is nearly the same as the separation energy of the
$\Upsilon(2S)$ and the $\Upsilon(1S)$ states. This will be important for
our studies of Heavy Quark Hybrids, but turns out to be misleading.

\subsection{The  $\rho-\pi$ Puzzle}

  The  $\rho-\pi$ puzzle for $c\bar{c}$ $1^{--}$ states is based on the
two diagrams for PQCD and electromagnetic decay of such states, shown in 
Fig. 2.
\begin{figure}[ht]
\begin{center}
\epsfig{file=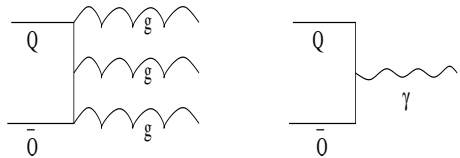,height=2cm,width=6cm}
\caption{Perturbative QCD and em diagrams for $Q \bar{Q}(1^{--})$ Decays}
\label{Fig.2}
\end{center}
\end{figure}
\noindent
By taking ratios the wave functions at the origin cancel, and this 
predicts the ratio of branching ratios for $c\bar{c}$ decays into hadrons (h)

\beq
\label{1}
  R&=&\frac{B(\Psi'(2S)\rightarrow h)}{B(J/\Psi(1S)\rightarrow h)}
\;=\;\frac{B(\Psi'(2S)\rightarrow e^+e^-)}{B(J/\Psi(1S)\rightarrow 
e^+e^-)} \nonumber \\
 &\simeq& 0.12  \; ,
\eeq
the famous 12\% rule. 

The $\rho-\pi$ puzzle: The $\Psi'(2S)$ to $J/\Psi$ 
ratios for $\rho-\pi$ and other hadron decays are more than an order of 
magnitude smaller than predicted by Eq({\ref{1}).   
Many theorists have tried and failed to explain this 
puzzle. See Ref\cite{cb98} for a review. This and more recent attempts at 
a solution are also discussed in Ref\cite{kpr08}, and all agree that 
previous work has not produced a solution for this puzzle.

\subsection{The sigma Decays of $\Upsilon(nS)$ States Puzzle}

  The puzzle of sigma decays of $b\bar{b}$ $1^{--}$ ($\Upsilon(nS)$) states 
is given by the following. The sigma is a low-energy broad two-pion scalar 
resonance. Experiments on $\Upsilon(nS)$ states find\cite{hv06}
\vspace{3mm}

$\Upsilon(2S) \rightarrow \Upsilon(1S) + 2\pi$ has a large branching ratio, 
but no $\sigma$
\vspace{3mm}

$\Upsilon(3s) \rightarrow \Upsilon(1S) + 2\pi$ has a large branching ratio to 
$\sigma$
\vspace{3mm}

\hspace{2cm}  $\Delta n = 2$, emit $\sigma$ 
\vspace{3mm}

 \hspace{2cm}  $\Delta n \neq 2$, no $\sigma$ emitted.
\vspace{5mm}

This is the Vogel $\Delta n=2$ puzzle, which cannot be understood using 
perturbative QCD, as expected for heavy bottomium states.

\subsection{The Octet Model for RHIC and Hybrids}

   The major goal of modern RHIC (Relativistic Heavy Ion Collision) 
experiments is to produce and study the quark-gluon plasma (QGP) which existed 
in the early universe before the QCD phase trasition, about $10^{-5}$ seconds 
after the big bang. One important signal of this QGP is the production 
of heavy quark (charmonium and upsilon) states via $q \bar{q}$ interactions 
in the early universe. The most natural mechanism is $q \bar{q} \rightarrow 
g \rightarrow Q \bar{Q}$, in which an octet $q \bar{q}$  produces an octet 
$Q \bar{Q}$, which is just PQCD, followed by the nonperturbative (NPQCD) 
process in which the octet $Q \bar{Q}$ becomes a singlet $Q \bar{Q}$ with 
the emission of a gluon (or other color octet). This is depicted in  Fig. 3:

\begin{figure}[ht]
\begin{center}
\epsfig{file=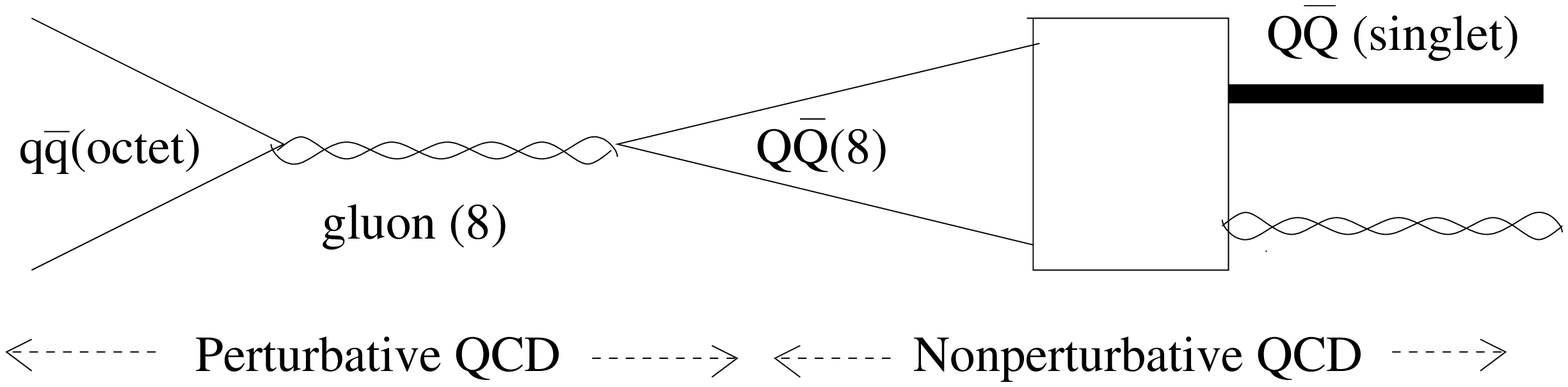,height=3cm,width=6cm}
\caption{Octet model for production of heavy quark mesons}
\label{Fig.3}
\end{center}
\end{figure} 

The nonperturbative matrix elements for the transition from the color 
octet $<Q \bar{Q}(8)|$ state to a color singlet $\Psi$ state, 
$<0|\mathcal{O}_8^{\Psi}|0>$ in the notation of Ref\cite{cl}, have been
determined by fits to experiments using the octet model\cite{cl}. As we shall
see, our determination of mixed heavy quark and heavy quark hybrid mesons
will provide a mechanism for predicting these NPQCD matrix elements.

\section{Mixed Heavy Quark Hybrid Heavy Quark $1^{--}$ States and QCD Sum 
Rules}

In this section we review the method of QCD sum rules, review our previous 
application of this method to attempt to find the lowest energy hybrid 
charmonium $1^{--}$ state, and present our new application of the QCD sum 
rule method to find the lowest energy mixed charmonium and upsilon states 
with hybrids.

\subsection{Method of QCD Sum Rules}

   The starting point of the method of QCD sum rules\cite{sz79} for finding 
the mass of a state A is the correlator,
 
\beq
\label{2}
       \Pi^A(x) &=&  \langle | T[J_A(x) J_A(0)]|\rangle \; ,
\eeq
with $| \rangle$ the vacuum state and
the current $J_A(x)$ creating the states with quantum numbers A:
\beq
\label{3}
     J_A(x)| \rangle &=& c_A |A \rangle + \sum_n c_n |n; A  \rangle  \; ,
\eeq
where $ |A \rangle$ is the lowest energy state with quantum numbers A,
and the states $|n; A  \rangle$ are higher energy states with the A quantum
numbers, which we refer to as the continuum.     

   The QCD sum rule is obtained by evaluating $\Pi^A$ in two ways. First, 
after a Fourier transform to momentum space, a dispersion relation gives the 
left-hand side (lhs) of the sum rule:
\beq
\label{4}
 \Pi(q)^A_{\rm{lhs}} &=&  \frac{\rm{Im}\Pi^A(M_A)}
{\pi(M_A^2-q^2)}+\int_{s_o}^\infty ds \frac{\rm{Im}\Pi^A(s)}
{\pi(s-q^2)}
\eeq
where $M_A$ is the mass of the state $A$ (assuming zero width)
and $s_o$ is the start of the continuum--a parameter to be determined.
The imaginary part of $\Pi^A(s)$, with the term for the state we are
seeking shown as a pole (corresponding to a $\delta(s-M_A^2)$ term in 
$\rm{Im}\Pi$), and the 
higher-lying states produced by $J_A$ shown as the continuum, is 
illustrated in Fig. 4:
\begin{figure}[ht]
\begin{center}
\epsfig{file=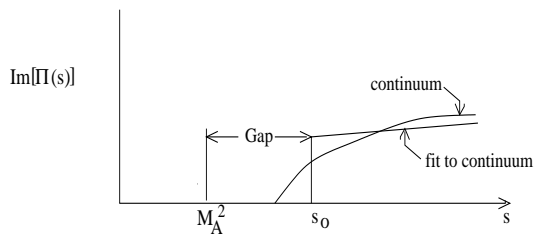,height=3cm,width=7cm}
\caption{QCD sum rule study of a state A with mass M$_A$ (no width)}
\label{Fig. 4}
\end{center}
\end{figure}

Next $ \Pi^A(q)$ is evaluated by an operator product expansion
(O.P.E.), giving the right-hand side (rhs) of the sum rule
\beq
\label{5}
  \Pi(q)_{\rm{rhs}}^A &=& \sum_k c_k(q) \langle 0|{\cal O}_k|0\rangle
 \; ,
\eeq
where $c_k(q)$ are the Wilson coefficients and $\langle 0|{\cal O}_k|0\rangle$
are gauge invariant operators constructed from quark and gluon fields,
with increasing $k$ corresponding to increasing dimension of ${\cal O}_k$.
It is important to note that the Wilson coefficients, $c_k(q)$ obey
renormalization group equations\cite{rry85}

  After a Borel transform, ${\mathcal B}$, in which the q variable is 
replaced by the Borel mass, $M_B$, 
\beq
\label{6}
       \mathcal{B}= \lim_{q^2,n\rightarrow \infty}\frac{1}{(n-1)!}
(q^2)^n(-\frac{d}{d q^2})^n\bigg\vert_{q^2/n = M_B^2} \; .
\eeq
the final QCD sum rule, ${\mathcal B} \Pi_A(q)(LHS) = 
{\mathcal B} \Pi_A(q)(RHS)$, has the form
\beq
\label{7}
    && \frac{1}{\pi} e^{-M_A^2/M_B^2}
+ {\cal B} \int_{s_o}^\infty \frac{Im[\Pi_A(s)]}{\pi(s-q^2)} ds \nonumber \\
     &=& {\cal B} \sum_k c_k^A(q) <0|{\cal O}_k|0> \; .
\eeq

 This sum rule and tricks are used to find $M_A$, which should vary little
with $M_B$. A gap between $M_A^2$ and $s_o$ is needed for accuracy. If the 
gap is too large, the solution is unphysical, which is important for our 
present work, as we discuss below.

\subsection{Hybrid charmonium}

  Here we give a brief review of the calculation of the correlator and
the results for the QCD sum rule for a pure hybrid charmonium $1^{--}$
state, which could possibly be the $\Psi'(2S)$. The current $J_{HH}$ (which 
we called $J_H$ in Ref\cite{kpr08}) for a heavy quark hybrid meson with 
$J^{PC} =1^{--}$ is
\beq
\label{8}
         J^\mu_{HH} &=&  \bar{\Psi}\Gamma_\nu G^{\mu\nu} \Psi \; ,
\eeq
where $\Psi$ is the heavy quark field, $\Gamma_\nu = C \gamma_\nu$,
$\gamma_\nu$ is the usual Dirac matrix, C is the charge conjugation operator,
and the gluon color field is
\beq
\label{9}
         G^{\mu\nu}&=& \sum_{a=1}^8 \frac{\lambda_a}{2} G_a^{\mu\nu}
\; ,
\eeq
with $\lambda_a$ the SU(3) generator ($Tr[\lambda_a \lambda_b]
= 2 \delta_{ab}$). The correlator
\beq
\label{10}
   \Pi_{HH}^{\mu \nu}(x) &=& <0|T[J_{HH}^\mu(x) J_{HH}^\nu(0)]|0> \; ,
\eeq
after a Fourier transform,
was evaluated using the leading two operators in the operator product 
expansion, shown in Figs. 5 and 6. The scalar correlator 
$\Pi^S$ is defined by $\Pi^{\mu \nu}(p)=(p_\mu p_\nu/p^2-g^{\mu \nu} )\Pi^V(p)
+ (p_\mu p_\nu/p^2) \Pi^S(p)$.

\begin{figure}[ht]
\begin{center}
\epsfig{file=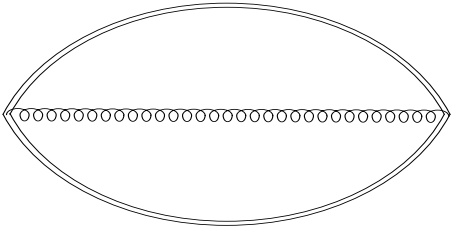,height=3cm,width=6cm}
\caption{ Lowest-order term in Sum Rule}
\label{Fig.5}
\end{center}
\end{figure}

\begin{figure}[ht]
\begin{center}
\epsfig{file=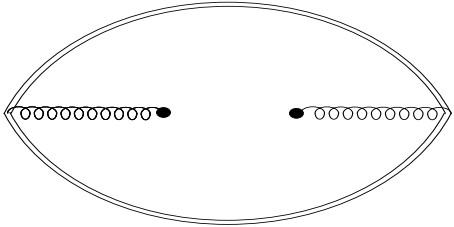,height=3cm,width=6cm}
\caption{Gluon condensate term in Sum Rule}
\label{Fig.6}
\end{center}
\end{figure}

   The leading term in the OPE for $\Pi_{HH}^S$, $\Pi_{1HH}^S$, corresponds
to the diagram in Fig. 5. It is quite complicated, and the result is 
given in Ref\cite{kpr08}. After the Borel transform,$\Pi_{1HH}^S(M_B)$ is
given in the Appendix, Eq(\ref{27}). 

    The second order term, corresponding to the operator with a gluon
condensate shown in Fig. 6, has a scalar part $\Pi_{2HH}^S(M_B)$ given
in the Appendix, Eq(\ref{28}).

    In Ref\cite{kpr08} a solution to the QCD Sum rule was found for a
charmonium hybrid state at the mass of $\Psi'(2S)$, from which one
would at first conclude that the $\Psi'(2S)$ is a pure hybrid $1^{--}$
meson. However, in order to satisfy the critrion that the solution is
almost independent of the Borel mass a value of $s_o= 60.0 Gev^2$ was
needed. This would imply that the next excited state was 7 to 8 GeV,
which is not consistent with the first state at only 3.66 GeV. Note that
lattice QCD calculations found the first charmonium hybrid at about one
GeV higher than our solution\cite{lm02,chen08}, which is also consistent the 
$\Psi'(2S)$ not being a pure hybrid. 

   This result, as well as the heavy quark puzzles and RHIC experiments
discussed in Sec II, were the main motivation for the present work, in 
which we seek a solution for a mixed charmonium and hybrid charmonium state.

\subsection{Mixed charmonium-Hybrid charmonium States}

  Recognizing that there is strong mixing between a heavy quark meson and
a hybrid heavy quark meson with the same quantum numbers (as shown below), 
and that the fact that our pure hybrid charmonium solution was not a physical 
state, we now attempt to find the lowest $J^{PC}=1^{--}$ charmonium state 
with a sizable admixture of a charmonium meson and a hybrid charmonium meson.
An appropriate mixed vector ($J^{PC}=1^{--}$) charmonium, hybrid charmonium 
current to use in QCD Sum Rules is
\beq
\label{11}
        J^\mu &=& b J_H^\mu + \sqrt{1-b^2} J_{HH}^\mu 
\eeq
with
\beq
\label{12}
          J_H^\mu &=& \bar{q}_c^a \gamma^\mu q_c^a  \; ,
\eeq
where  $J_H^\mu$ is the standard current for a $1^{--}$ charmonium state, and 
$J_{HH}^\mu$ is the heavy charmonium hybrid current given above in Eqs(\ref{8},
\ref{9}).

Therefore the correlator for the mixed state:
\beq
\label{13} 
   \Pi_{H-HH}^{\mu\nu}(x) &=& <0|T[J^\mu(x) J^\nu(0)]|0>
\eeq
is 
\beq
\label{14}
   \Pi_{H-HH}^{\mu\nu}(x) &=& b^2  \Pi_{H}^{\mu\nu}(x) + (1-b^2)
\Pi_{HH}^{\mu\nu}(x) \nonumber \\
  && +2b\sqrt{1-b^2}\Pi_{HHH}^{\mu\nu}(x) \\
       \Pi_{H}^{\mu\nu}(x)&=& <0|T[J_H^\mu(x) J_H^\nu(0)]|0> \nonumber \\
   \Pi_{HH}^{\mu\nu}(x)&=& <0|T[J_{HH}^\mu(x) J_{HH}^\nu(0)]|0> \nonumber \\
   \Pi_{HHH}^{\mu\nu}(x)&=& <0|T[J_H^\mu(x) J_{HH}^\nu(0)]|0> \nonumber \; .
\eeq
For heavy quarks the gluon condensate is proportional to the quark
condensate, and the renormalization group equations for the Wilson
coefficients of the operator product expansions of $\Pi_{H}$, $\Pi_{HH}$,
and $\Pi_{HHH}$ are similar for the terms considered here\cite{rry85}. 
  
The heavy hybrid correlator, $\Pi_{HH}^{\mu\nu}(q^2)$ was presented in the 
previous section. The operator product expansion for the standard heavy
quark correlator, $\Pi_{H}^{\mu\nu}(q^2)$ (RHS) is given by the diagrams
shown in Fig. 7.
\begin{figure}[ht]
\begin{center}
\epsfig{file=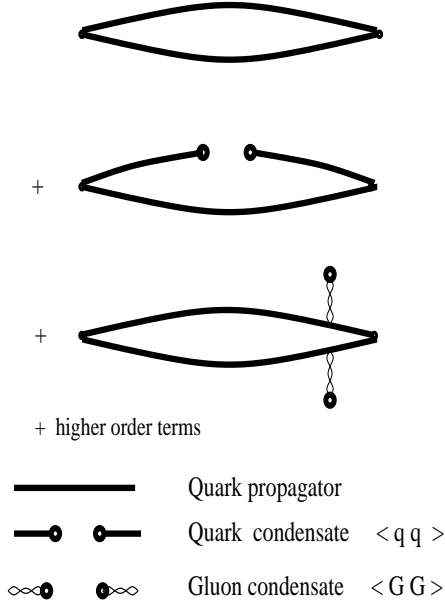,height=8cm,width=6cm}
\caption{Heavy quark meson diagrams}
\label{Fig.7}
\end{center}
\end{figure}

    The leading term for the quark correlator in momentum space,
with $M_C$ the charm quark mass, is
\beq
\label{15}
  \Pi_{H1}^{\mu \nu }(p)& =& g_v^2\int \frac{d^4 k}{(2 \pi)^4} Tr[S(k) 
\Gamma_5^\mu S(p-k)\Gamma_5^{\nu T}] \nonumber \\
          S(k) &=& \frac{\not\!k + M_C}{k^2-M_C^2} \\
        \Gamma_5^\mu &=& \gamma^\mu \gamma_5 \nonumber \; .
\eeq

   Noting that the charmonium quark condensate is very small, and that the
gluon condensate term and all higher-dimensional terms are also small,
$\Pi_{H1}^{\mu \nu }(p)$ dominates the heavy quark correlator, 
$\Pi_{H}^{\mu \nu }(p)$.  Carrying out the momentum integral in Eq(\ref{15}),
and extracting the scalar correlator we find
\beq
\label{16}
       \Pi_H^S(p)& =& i  \frac{3 g_v^2}{(4 \pi)^2}
\int_o^1 d\alpha \frac{6 p^4 -23 p^2 M_C^2}{(\alpha-\alpha^2) p^2-M_C^2}
\eeq
Carrying out the Borel transform we find 
\beq
\label{17}
   \Pi_{H}^S(M_B) &=& \frac{3}{2 \pi^2} M_C^4 exp^{-2 z}[\frac{13}{4}K_o(2 z)
+\frac{1}{2} K_1(2 z) \nonumber \\
   &&+3 K_2(2 z)]  \; ,
\eeq 
with $z=M_C^2/M_B^2$.

   Finally, for the $\Pi_{HHH}^{\mu\nu}$ term, the dominant diagram is shown
in Fig. 8, in which the gluon from the $J_{HH}$ operator is coupled to a
quark, leading to the $J_H$ operator. This is essentially the perturbative 
plus nonperturbative H-HH matrix element without condensates.
\begin{figure}[ht]
\begin{center}
\epsfig{file=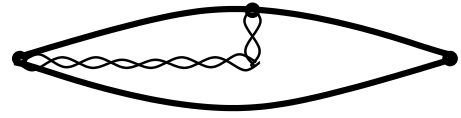,height=1.5cm,width=6cm}
\caption{Meson-hybrid meson lowest order diagram}
\label{Fig. 8}
\end{center}
\end{figure}

  Using the external field method, the leading term of  $\Pi_{HHH}^{\mu\nu}$,
corresponding to Fig. 8, is
\beq
\label{18}
    \Pi_{HHH1}^{\mu \nu }(p)& =& -i \frac{g_v^2}{4}\int \frac{d^4 k}{(2 \pi)^4}
Tr[\frac{[\sigma_{\kappa \delta},(\not\!k + M_C)]_{+}}{(k^2-M_C^2)} 
\nonumber \\
  && \frac{C \gamma_\lambda (\not\!p -\not\!k + M_C)(C \gamma_\mu)^T}
{(p-k)^2-M_C^2}] \nonumber \\
  && Tr[G^{\nu \lambda}(0) G^{\kappa \delta}(0)] \; .
\eeq

   After a Borel transform and extracting the scalar component of 
$\Pi_{HHH1}^{\mu \nu }$, one finds that 
\beq
\label{19}
       \Pi_{HHH1}^S(M_B) &\simeq& \pi^2  \Pi_{H}^S(M_B)  \; .      
\eeq

Therefore thr right-hand side of our scalar correlator is
\beq
\label{20}
   \Pi_{H-HH}^S(M_B)_{rhs} &=& (b^2 + 2 \pi^2 b \sqrt{1.-b^2}) \Pi_{H}^S(M_B)
\nonumber \\
   &&+(1.-b^2)\Pi_{HH}^S(M_B) \; .
\eeq
The left hand side of the sum rule has the usual form (see Eq(7))
\beq
\label{21}
  &&  \Pi_{H-HH}^S(M_B)_{lhs} = F e^{-M_{H-HH}^2/M_B^2} \\
   && +e^{-s_o/M_B^2} (K0 + K1 M_B^2 + K2 M_B^4 + K3 M_B^6)
\nonumber \; ,
\eeq
with $s_o, K0, K1, K2, K3$ parameters used to fit the continuum.
Note that the meson and hybrid meson states associated with the H and HH
operators are normalized independently, and the operators have different 
dimensions. We renormalize by calculating $NHH = \int d M_B 
\Pi_{H}^S(M_B)/ \int d M_B \Pi_{HH}^S(M_B)$. Henceforth, for $\Pi_{HH}^S(M_B)$
we use $NHH \times \Pi_{HH}^S(M_B)$.

As in Ref\cite{kpr08}, we obtain the expression for the mass of the mixed
heavy meson-hybrid heavy meson by taking the ratio of the derivative of the
sum rule with res[ect to $1/M_B^2$ to the sum rule, giving 
\beq
\label{22}
 && M_{H-HH}^2 = \{[s_o(K0+K1 M_B^2+K2 M_B^4+K3 M_B^6) \nonumber \\
&& +K1 M_b^4 +2 K2 M_B^6+3 K3 M_B^8] e^{-\frac{s_o}{M_B^2}} \nonumber \\
 &&+ \partial_{1/M_B^2} \Pi_{H-HH}^S \} 
  \times \{(K0+K1 M_B^2 +K2 M_B^4  \nonumber \\
 && +K3M_B^6) e^{-\frac{s_o}{M_B^2}}-\Pi_{H_HH}^S \}^{-1} \; .
\eeq
 
   A key parameter in our numerical fits is the value of b. The solution
for b=-0.7 was most successful in fitting the criteria for finding the
mixed hybrid state using QCD sum rules. The range of b for which a 
satisfactory solution is obtained is b=$-0.7 \pm 0.1$, with the result for
b=-.7 shown in Fig. 9.
\begin{figure}[ht]
\begin{center}
\epsfig{file=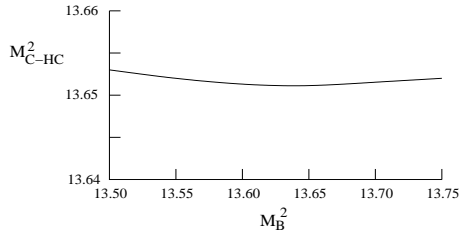,height=3.0cm,width=6cm}
\caption{Mixed charmonium-hybrid charmonium mass = 3.69GeV}
\label{Fig. 9}
\end{center}
\end{figure}

  We find the mass of the lowest-energy mixed charmonium-hybrid charmonium
to be about the energy of the $\Psi'(2S)$ state, 3.69 GeV, with
$s_o$=20 GeV$^2$, b= -0.7 $\Rightarrow$ 50-50 per cent charmonium-hybrid 
charmonium. It satisfies the criteria for about a fifteen per cent accuracy. 
The values of the other parameters are $K0=-15.9, K1=0.224, K2=-0.00015,
K3=0.00009$. The only solutions satisfying the sum rule criteria are
those with the value of b about $-.7 \pm.1$, so that we find the state to be 
about a 50-50 per cent meson-hybrid meson. As we shall see, this gives a 
solution to the $\rho-\pi$ puzzle.

\subsection{Mixed upsilon-hybrid upsilon states}

   The calculation of the mixed upsilon-Hybrid upsilon meson mass is the
same as that of the mixed charmonium-hybrid charmonium mass using QCD sum
rules, except the charm quark mass (which we took as $M_C^2$ = 1.8 Gev$^2$)
is replaced by the bottom quark mass (which we take as $M_b^2$ = 25.0 Gev$^2$).
In fact, the QCD sum rule method is more accurate for the calculation of
upsilon states, since the bottom quark condensate is much smaller that the 
charm quark condensate, and the operator product expansion converges
faster. 

   Since we found that the $\Psi'(2S)$ is a mixed charmonium state (see 
previous subsection) and as we noted earlier the separation in energy between 
the $\psi'(2S)$ and $J/\psi(1S)$ states is nearly the same as the separation 
energy of the $\Upsilon(2S)$ and the $\Upsilon$(1S) states (see Fig 1), we
would expect that the $\Upsilon(2S)$ is a 50-50 mixture of upsilon and
hybrid upsilon. This in not our result, as we shall now see.

   From the QCD sum rule one obtains the expression given in Eq.(\ref{22}),
except the charm quark mass is replaced by the bottom quark mass in the
expressions for the right-hand side of the correlator. The parameters 
$s_o, K0, K1, K2, K3$ are chosen to fit the continuum, and the mixing
parameter b is also chosen to give a solution in which mixed upsilon state 
mass is almost independent of the Borel mass. The result is shown in Fig. 10.

\begin{figure}[ht]
\begin{center}
\epsfig{file=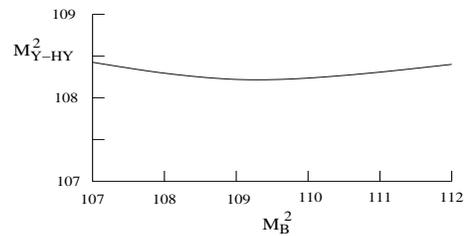,height=3.0cm,width=6cm}
\caption{Mixed upsilon-hybrid upsilon mass = 10.4 GeV}
\label{Fig. 10.}
\end{center}
\end{figure}

  We find the energy of the lowest mixed upsilon meson and hybrid upsilon
meson state to be at 10.4 GeV, approximately the energy of the $\Upsilon$(3S)
state (see Fig. 1.). The parameters are $s_o$ = 120 GeV$^2$, $K0=-50,000.,
K1=70500., K2=-605., K3=-0.5165$, and b$\simeq$ -0.7 for a good solution. Thus
we predict that the $\Upsilon$(3S) state is a 50-50 percent admixture of
an upsilon meson and a hybrid upsilon meson. As we shall now see, from this
we have obtained a solution to the Vogel $\Delta n=2$ puzzle.

\section{Mixed Meson-Hybrid Meson, Heavy Qyark Decay Puzzles, and Octet Model} 

  In this section we show that the solutions for the mixed nature of the
charmonium $\psi'(2S)$ and bottomonium $\Upsilon(3S)$ states provide
explanations for the decay puzzles and a basis for the calculation of
the nonperturbative matrix elements needed for the octet model used 
in RHIC calculations.

 \subsection{The $\rho-\pi$ Puzzle}

   First note that the matrix element $<\pi \rho|O|\psi'(c\bar{c},2S)>$ 
for  $\rho-\pi$ decay of $|c\bar{c}(2S)>$ is given by the PQCD diagram
shown in Fig. 11
\begin{figure}[ht]
\begin{center}
\epsfig{file=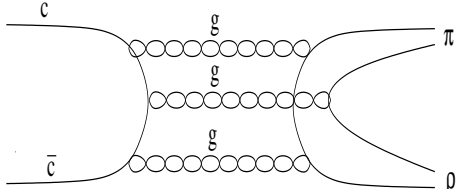,height=2.5cm,width=6cm}
\caption{PQCD diagram for charmonium decay into a $\pi$ and a $\rho$}
\label{Fig. 11}
\end{center}
\end{figure}

Next, the hybrid decay matrix element \\
 $<\pi \rho|O'|\psi'(c\bar{c}g,2S)>$ is given by the PQCD diagram shown
in Fig. 12.
\begin{figure}[ht]
\begin{center}
\epsfig{file=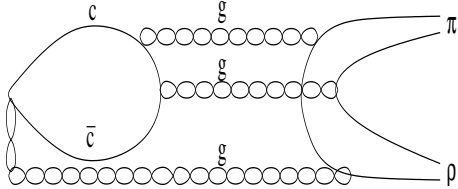,height=2.5cm,width=6cm}
\caption{ PQCD diagram for hybrid charmonium decay into a $\pi$ and a $\rho$}
\label{Fig. 12}
\end{center}
\end{figure}

As one can see from the diagrams, these matrix elements are almost equal in 
magnitude. Since we find that $|\Psi'(2S)>\simeq -0.7|c\bar{c}(2S)>
+0.7|c\bar{c}g(2S)>$, so the charmonium and hybrid charmonium approximately
cancel, we obtain for all 2 hadron decays, including $\rho+\pi$ decay,
\beq
\label{23}
  R&=&\frac{B(\Psi'(2S)\rightarrow \rho+\pi)}{B(J/\Psi(1S)
\rightarrow \rho+\pi)}<<\; 0.12 \; ,
\eeq
which is our proposed solution to the $\rho-\pi$ puzzle.

\subsection{$\sigma$ Decays of $\Upsilon(nS)$ States Puzzle}

   The solution to the Vogel $\Delta n=2$ puzzle is based on the aplication
of the glueball/sigma model, based on the study of scalar mesons and scalar
glueballs\cite{lsk97,kj01}, which was motivated by the BES analysis of
glueball decay\cite{bes97}, and our solution for
the lowest mixed state to be the $\Upsilon(3S)$ state. The glueball/sigma model
has been used for prediction of sigma production from glue created in
hadron-hadron collisions\cite{kms05} and the decay of hybrid 
baryons\cite{kl99}, which is closely related to the puzzle of sigma decays 
from upsilon states. The key is the glueball-meson coupling 
theorem\cite{nov80}
\beq
\label{24}
     \int dx T[J^G(x)J^m(0)]&\simeq& -\frac{32}{9}<\bar{q}q> \; ,
\eeq
where $<\bar{q}q> \equiv {\rm Quark\;Condensate}$,
which is depicted in Fig. 13.
\begin{figure}[ht]
\begin{center}
\epsfig{file=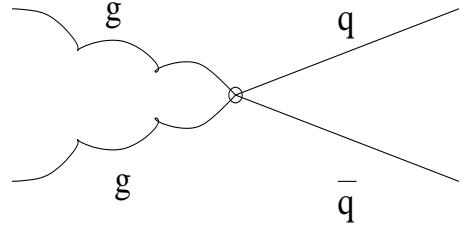,height=3.0cm,width=6cm}
\caption{Glueball-meson coupling}
\label{Fig.13}
\end{center}
\end{figure}

From this one can calculate the matrix element for sigma decay from a
hybrid meson, using the diagram shown in Fig. 14. 

\begin{figure}[ht]
\begin{center}
\epsfig{file=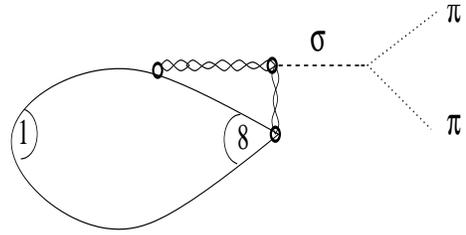,height=3.0cm,width=6cm}
\caption{Sigma decay of a hybrid meson}
\label{Fig.14}
\end{center}
\end{figure}

Just as scalar glueballs, such as the f$_o$(1500), decay mainly into sigmas, 
the hybrid component of the $\Upsilon$(3S) has a strong $\sigma$ decay branch, 
while we predict that the $\Upsilon$(2S) two-pion decay to the $\Upsilon$(1S) 
would have a very small $\sigma$ decay branch. Therefore, our solution for 
the $\Upsilon$(3S) to be a mixed $b \bar{b}$-$b \bar{b} g$ provides a 
solution to the Vogel $\Delta n=2$ puzzle. Since our states are not 
normalized we cannot calculate the numerical value of the cross section, 
a subject for future research.

\subsection{Octet Model for RHIC} 

   As discussed in section II, the octet model, depicted in Figure 3.,
is the dominant mechanism for production of heavy quark states from
a quark-gluon plasma produced via RHIC. Let us consider the collision
of a nucleus A, e.g., lead or gold nucleus, with a similar nucleus.
The differential cross section for the production of a 
charmonium state in a A-A collision in the color octet model is
\beq
\label{25}
&&\frac{d \sigma}{d p_T}[pp\rightarrow \psi(c \bar{c})]=
\int f_{q/A} f_{q/A}  \frac{d \sigma}{d t}[qq\rightarrow C\bar{C}(8)
 \nonumber \\
&&\rightarrow \psi(c \bar{c})] \nonumber \\
&&\frac{d \sigma}{d t}[qq\rightarrow C\bar{C}(8) \rightarrow \psi(c \bar{c})] 
=[{\rm perturbative\;QCD}] \nonumber \\
&& \times <0|\mathcal{O}_8^{\psi_Q}|0> \; ,
\eeq
where $f_{q/A}$ is the momentum fraction carried by a quark in the nucleus A,
and $<0|\mathcal{O}_8^{\psi_Q}|0>$ is the NPQCD color octet matrix 
element. In previous applications of 
the model, the nonperturbative octet-singlet matrix element was taken
from fits to other experiments\cite{nlc03,cln04}. We, however, can 
determine the NPQCD matrix elements for an octet quarkonium
pair to emit a gluon (octet) and leave a physical singlet quarkonium state,
which is given by $\Pi_{8,1}$. This is shown in Fig. 15.

\begin{figure}[ht]
\begin{center}
\epsfig{file=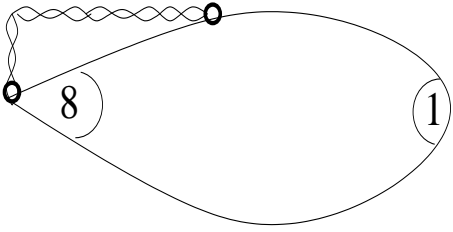,height=3cm,width=6cm}
\caption{$\Pi_{HHH}=\Pi_{8,1} \propto$ color 0ctet-color singlet matrix 
element}
\label{Fig.15}
\end{center}
\end{figure}

We have seen how to evaluate this diagram, but the normalization of the
states must be carried out to make a numerical estimate. We can, however,
estimate ratios of matrix elements, to predict ratios of quarkonium 
production. As an example, from a table in Cho-Leibovich\cite{cl}
\beq
\label{26}
 &&<0|\mathcal{O}_8^{J/ \psi}(1S)|0> = 1.2 \times10^{-2} GeV^3 \nonumber \\
 &&<0|\mathcal{O}_8^{\psi'}(2S)|0> = 0.73 \times10^{-2} GeV^3  {\rm \;or\;}
\nonumber \\
&&R_C=\frac{<0|\mathcal{O}_8^{\psi'}(2S)|0>}{<0|\mathcal{O}_8^{J/\psi}
(1S)|0>} \simeq \;0.6 \; .
\eeq

Since in our model of the $\psi'(2S)$ state the $c\bar{c}(1)$ component
should dominate, the parameter b$\simeq -.7$  gives a rough estimate
of this ratio, in agreement with the Cho-Leibovich phenomenological fit.

\section{Conclusions}

  In summary, we find that the $\psi'$(2S) is approximately 50\% charmonium
and 50\% hybrid charmonium; and the $\Upsilon$(3S) is approximately 50\% 
bottomium and 50\% hybrid bottomium.  This solves the $\rho-\pi$ problem 
for charmonium decays, and the Vogel $\Delta n=2$ puzzle for sigma decays
of upsilon states.

  From the correlator corresponding to the mixed heavy meson and heavy hybrid
meson current, the color octet-singlet matrix element can be obtained. This
nonperturbative matrix element can be used for studies of the production
of heavy quark states in RHIC experiments, using the octet model. It can
also be used with the sigma/glueball model to predict the cross sections 
for sigma production fron heavy quark state decays. Since the states
used in the QCD sum rule method are not normalized, these numerical
estimates cannot be made at the present time

   In the near future we plan to extend our calculation, so that numerical
predictions of heavy quark decays and RHIC production of heavy quark
states can be made. This will include possible tests of RHIC quarkonium
production via sigma decays.

\section{APPENDIX}

   In Ref\cite{kpr08} we found that $ \Pi_{1HH}^S(M_B)$, the Borel transform
of the scalar term of the the main diagram for the HH correlator, shown in 
Fig. 5, is

\beq
\label{A1}
   \Pi_{1HH}^S(M_B) &=&  -g_v^2\frac{1}{2(4\pi)^2}M_Q^4 \int_0^\infty d\delta
e^{-2 \frac{M_Q^2}{M_B^2}(1+\delta)}\nonumber \\
&&\{[-310 \frac{\delta}{1+\delta}+638 \delta -656 (1+\delta)] \nonumber \\
 &&K_3(2\frac{M_Q^2}{M_B^2}(1+\delta)) + [-1892 \frac{\delta}{1+\delta}+
\nonumber \\
&& 606 \delta -1968 (1+\delta) -32(4 \delta-3 \frac{\delta^2}{1+\delta}
\nonumber \\
&& + \frac{\delta^3}{3(1+\delta)^2})] K_2(2\frac{M_Q^2}{M_B^2}(1+\delta))
\nonumber \\
 && + [-4778 \frac{\delta}{1+\delta}+9442 \delta
-4920 (1+\delta) \nonumber \\
 &&-128(4 \delta-3 \frac{\delta^2}{1+\delta}
 + \frac{\delta^3}{3(1+\delta)^2})]\nonumber \\
&& K_1(2\frac{M_Q^2}{M_B^2}(1+\delta))+ [-1356 \frac{\delta}{1+\delta}
 \nonumber \\
 && +6284 \delta -3280 (1+\delta) \nonumber \\
 &&-96(4 \delta-3 \frac{\delta^2}{1+\delta}+ \frac{\delta^3}{3(1+\delta)^2})]
 \nonumber \\
&&  K_0(2\frac{M_Q^2}{M_B^2}(1+\delta))\}\nonumber \\
&&+{\rm multiple\;integrals} \; .
\eeq
The multiple integral terms in Eq.(\ref{27}) are small and are dropped. 
$M_Q$ is the charm quark mass for the charmonium calculations and the bottom 
quark mass for the upsilon calculations. We take $M_C^2$ = 1.8 GeV$^2$ and
$M_b^2$ = 25.0 GeV$^2$.
   The gluon condensate term is shown in Fig. 6. After the Borel transform
the scalar part of this term\cite{kpr08}, $ \Pi_{2HH}^S(M_B)$, is
\beq
\label{A2}
   \Pi_{2HH}^S(M_B) &=&  -ig_v^2\frac{3}{2(4\pi)^2}M_Q^4e^{-2 \frac{M_Q^2}
{M_B^2}}
\eeq
\vspace{-5mm}
\beq
 && [11 K_2(2\frac{M_Q^2}{M_B^2})+\frac{14}{3}K_1(2\frac{M_Q^2}{M_B^2})
 +18 K_0(2\frac{M_Q^2}{M_B^2})]  \nonumber\; .
\eeq

The $K_n$ are Bessel functions of imaginary argument, related to Hankel
functions by
\beq
\label{A3}
          K_n(x) &=& \frac{i \pi}{2} e^{-n\frac{i \pi}{2}} H^{(1)}_n(ix)
\eeq

\newpage

\large{{\bf Acknowledgments}}
\normalsize
\vspace{1mm}

  This work was supported in part by the NSF/INT grant number 0529828.

  The author thanks Drs. Diana Parno, Seamus Riordan, Ming Liu, and Pat
McGoughey; Professors Pengnian Shen and Wei-xing Ma, and other IHEP, Beijing 
colleagues, and Professors Roy Briere and Helmut Vogel for helpful discussions.
We thank Professor Y. Chen for discussions of lattice QCD in comparison to QCD 
sum rules for hybrid states.

\end{document}